\documentclass[twocolumn,amsmath,amssymb]{revtex4-1}
\usepackage{graphicx}
\usepackage{dcolumn}
\usepackage{bm}
\usepackage{color}

\newcommand{\rr}{\textbf{r}}
\newcommand{\xx}{\textbf{r}}

\newcommand\etadisp{\ensuremath{\eta_\textit{d}}}
\newcommand\epsilondisp{\ensuremath{\epsilon_\textit{d}}}
\newcommand\epsilonassoc{\ensuremath{\epsilon_\textit{a}}}
\newcommand\kappaassoc{\ensuremath{\kappa_\textit{a}}}
\newcommand\lambdadisp{\ensuremath{\lambda_\textit{d}}}
\newcommand\lscale{\ensuremath{s_d}}

\newcommand\hughesetal{Hughes \emph{et al.}}

\newcommand\hughesetalcite{Hughes \emph{et al.}~\cite{hughes2013classical}}

\begin{document}
\title{Improved association in a classical density functional theory
  for water}

\author{Eric J. Krebs}
\affiliation{Department of Physics, Oregon State University,
  Corvallis, OR 97331, USA}
\author{Jeff B. Schulte}
\affiliation{Department of Physics, Oregon State University,
  Corvallis, OR 97331, USA}
\author{David Roundy}
\affiliation{Department of Physics, Oregon State University,
  Corvallis, OR 97331, USA}

\begin{abstract}
We present a modification to our recently published SAFT-based classical
density functional theory for water.
We have recently developed and tested a functional for the averaged
radial distribution function at contact of the hard-sphere fluid
that is dramatically more accurate at interfaces than earlier
approximations.
We now incorporate this improved functional into the association term
of our free energy functional for water, improving its description of
hydrogen bonding.
We examine the effect of this improvement by studying two hard solutes: a
hard hydrophobic rod and a hard sphere.
The improved functional leads to a moderate change in the density
profile and a large decrease in the number of hydrogen bonds broken in
the vicinity of the solutes.
\end{abstract}
\maketitle

\section{Introduction}

Water, the universal solvent, is of critical practical importance, and
a continuum description of water is in high demand for a solvation
model.  A number of recent attempts to develop improved solvation
models for water have built on the approach of classical
density functional theory (DFT)~\cite{jeanmairet2013molecular,
  zhao2011molecular, zhao2011new, ramirez2005direct,
  ramirez2005density, levesque2012solvation, levesque2012scalar}.
Classical DFT is based on a description of a fluid written as a free
energy functional of the density distribution.  There are two general
approaches used to construct a classical DFT for water.  The first is
to choose a convenient functional form which is then fit to properties
of the bulk liquid at a given temperature and pressure
\cite{jeanmairet2013molecular, zhao2011molecular, zhao2011new,
  ramirez2005direct, ramirez2005density, levesque2012solvation,
  levesque2012scalar, lischner2010classical}.  Using this approach, it
is possible to construct a functional that reproduces the exact
second-order response function of the liquid under the fitted
conditions.  However, this class of functional will be less accurate
at other temperatures or pressures---and in the inhomogeneous
scenarios in which solvation models are applied.  The second approach
is to construct a functional by applying liquid-state theory to a
model system, and then fit the model to experimental data such as the
equation of state~\cite{hughes2013classical, clark2006developing,
  gloor2002saft, gloor2004accurate, gloor2007prediction, Jaqaman2004,
  chuev2006, fu2005vapor-liquid-dft,kiselev2006new,
  blas2001examination, sundararaman2012computationally}.

A widely used family of models used in the development of classical
density functionals is based on Statistical Associating Fluid Theory
(SAFT)~\cite{chapman1989saft}.  SAFT is a theory based on a model of
hard spheres with weak dispersion interactions and hydrogen-bonding
association sites, which has been used to accurately model the
equations of state of both pure fluids and mixtures over a wide range
of temperatures and pressures~\cite{muller2001molecular,
  tan2008recent}. The association contribution to the free energy uses
Wertheim's first-order thermodynamic perturbation theory to describe
an associating fluid as hard-spheres with strong associative
interactions at specific sites on the surface of each
sphere~\cite{wertheim1984fluidsI, wertheim1984fluidsII,
  wertheim1986fluidsIII, wertheim1986fluidsIV}.  These association
sites have an attractive interaction at contact, and rely on the
hard-sphere pair distribution function at contact $g_\sigma^\text{HS}$
in order to determine the extent of association.  While this function
is known for the homogeneous hard-sphere fluid, it must be
approximated for inhomogeneous systems, such as occur at liquid
interfaces.

In a recent paper, we examined the pair distribution function at
contact in various inhomogeneous
configurations~\cite{schulte2012using}.  We tested the accuracy of
existing approximations for the pair distribution function at
contact~\cite{yu2002fmt-dft-inhomogeneous-associating,
  gross2009density}, and derived a significantly improved
approximation for the averaged distribution function at contact.  In
this paper we apply this improved $g_\sigma^{HS}$ to the SAFT-based
classical density functional for water developed by Hughes \emph{et
  al.}~\cite{hughes2013classical}.  This functional was constructed to
reduce in the homogeneous limit to the 4-site optimal SAFT model for
water developed by Clark \emph{et al.}~\cite{clark2006developing}.
The DFT of \hughesetal\ uses the association free energy functional of
Yu and Wu~\cite{yu2002fmt-dft-inhomogeneous-associating}, which is
based on a $g_\sigma^{HS}$ that we have since found to be
inaccurate~\cite{schulte2012using}.  In this paper, we will examine
the result of using the improved functional for $g_\sigma^{HS}$
developed in \hughesetal\ to construct an association free energy
functional.

\section{Method}

The classical density functional for water of \hughesetal\ consists of
four terms:
\begin{align}
  F[n(\rr)] &= F_{\text{ideal}}[n(\rr)] + F_{\text{HS}}[n(\rr)]
  + F_{\text{disp}}[n(\rr)] + F_{\text{assoc}}[n(\rr)]
\end{align}
where $F_{\text{ideal}}$ is the ideal gas free energy and
$F_{\text{HS}}$ is the hard-sphere excess free energy, for which we
use the White Bear functional~\cite{roth2002whitebear}.
$F_{\text{disp}}$ is the free energy contribution due to the
square-well dispersion interaction; this term contains one empirical
parameter, $\lscale$, which is used to fit the surface tension of water near one
atmosphere.  Finally, $F_{\text{assoc}}$ is the free energy
contribution due to association, which is the term that we examine in
this paper.

\subsection{Dispersion}
The dispersion term in the free energy includes the van
der Waals attraction and any orientation-independent
interactions. Following \hughesetal, we use a dispersion term based on
the SAFT-VR approach\cite{gil-villegas-1997-SAFT-VR}, which has two
free parameters (taken from Clark~\emph{et
  al}\cite{clark2006developing}): an interaction energy $\epsilondisp$
and a length scale $\lambdadisp R$.

The SAFT-VR dispersion free energy has the form~\cite{gil-villegas-1997-SAFT-VR}
\begin{align}
  F_\text{disp}[n] &= \int \left(a_1(\xx) + \beta a_2(\xx)\right)n(\xx)d\xx
  \label{eq:dispersion-term}
\end{align}
where $a_1$ and $a_2$ are the first two terms in a high-temperature
perturbation expansion and $\beta=1/k_BT$.  The first term, $a_1$, is 
the mean-field dispersion interaction. The second term, $a_2$, describes the
effect of fluctuations resulting from compression of the fluid due
to the dispersion interaction itself, and is approximated
using the local compressibility approximation (LCA), which
assumes the energy fluctuation is simply related to the
compressibility of a hard-sphere reference fluid\cite{barker1976liquid}.

The form of $a_1$ and $a_2$ for SAFT-VR is given in
reference~\cite{gil-villegas-1997-SAFT-VR}, expressed in terms
of the packing fraction.  In order to apply this form to an
\emph{inhomogeneous} density distribution, we construct an effective local
packing fraction for dispersion $\etadisp$, given by a Gaussian
convolution of the density:
\begin{align}
  \etadisp(\xx) &= \frac{1}{6\sqrt{\pi} \lambdadisp^3\lscale^3}
  \int n(\xx')\exp\left({-\frac{|\xx-\xx'|^2}{2(2 \lambdadisp
      \lscale R)^2}}\right)d\xx'.\label{eq:packing-fraction}
\end{align}
This effective packing fraction is used throughout the dispersion
functional, and represents a packing fraction averaged over the
effective range of the dispersive interaction.
Eq.~\ref{eq:packing-fraction} contains an additional empirical
parameter $\lscale$ introduced by \hughesetal, which modifies the
length scale over which the dispersion interaction is correlated.

\begin{figure}
\begin{center}
\includegraphics[width=3.5in]{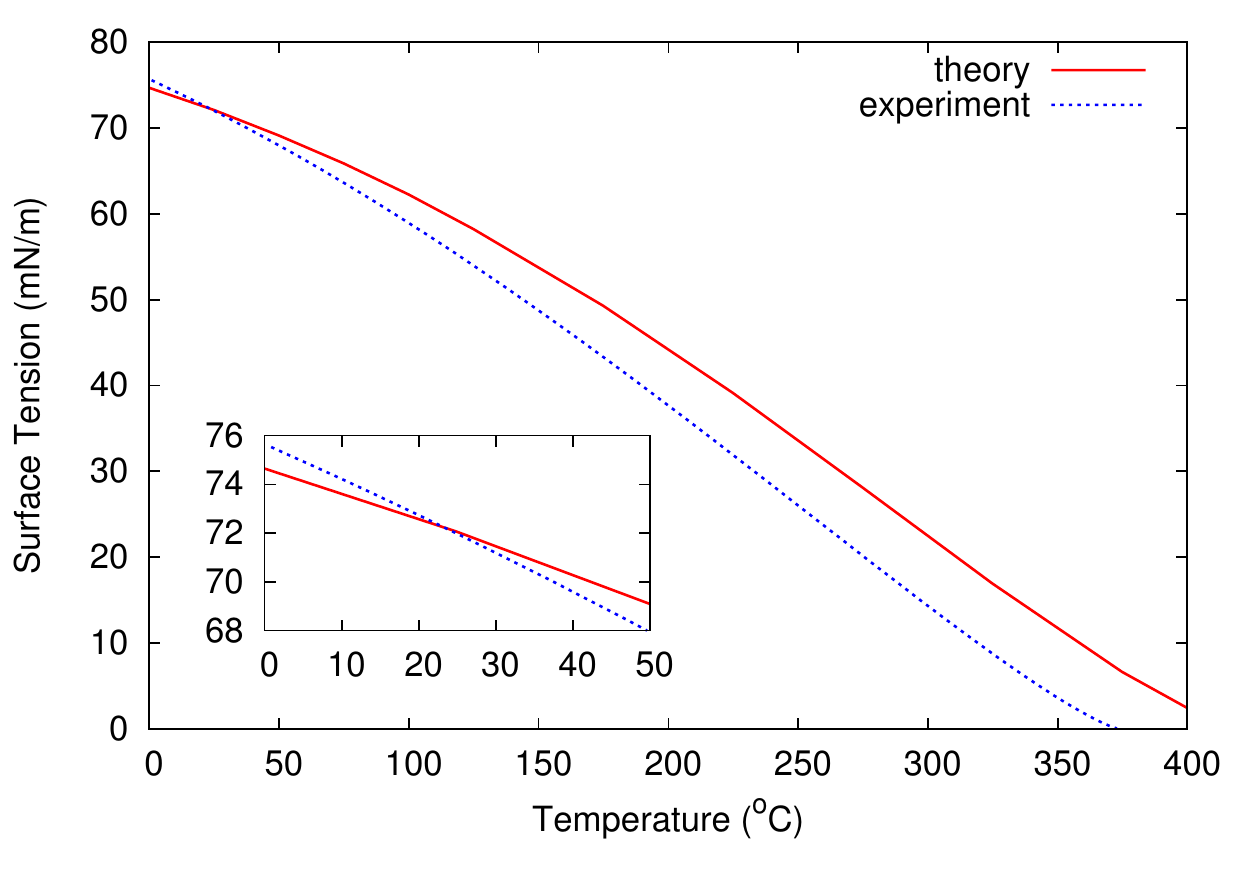}
\end{center}
\caption{Comparison of Surface tension versus temperature for
  theoretical and experimental data. The experimental data is taken
  from NIST~\cite{nistwater}.  The length-scaling parameter $\lscale$
  is fit so that the theoretical surface tension will match the
  experimental surface tension near room temperature.}
\label{fig:surface-tension}
\end{figure}

\subsection{Association}

The association free energy for our four-site model has the form:
\begin{align}
  F_\text{assoc}[n] &= k_BT \int n_\text{site}(\xx)
  \left(\ln X(\xx) - \frac{X(\xx)}{2} + \frac12\right) d\xx
\end{align}
where $n_\text{site}(\rr)$ is the density of bonding sites at
position~$\rr$:
\begin{align}
  n_\text{site}(\rr) &=
  \begin{cases}
    4 n(\rr) & \text{this work}\\
    4 n_0(\rr) \zeta(\rr) & \text{\hughesetalcite}
  \end{cases}
\end{align}
where the factor of four comes from the four hydrogen bond sites, the
fundamental measure $n_0(\rr)$ is the average density contacting point
$\rr$, and $\zeta(\xx)$ is a dimensionless measure of the density
inhomogeneity from Yu and
Wu~\cite{yu2002fmt-dft-inhomogeneous-associating}.  The functional
$X(\rr)$ is the fraction of association sites \emph{not}
hydrogen-bonded, which is determined for our 4-site model by the
quadratic equation
\begin{align}
  X(\xx) &= \frac{\sqrt{1 + 2n_\text{site}'(\rr)
      \kappaassoc g^\textit{SW}_\sigma(\xx)
  \left(e^{-\beta\epsilonassoc} - 1\right)} - 1}
  {n_\text{site}'(\rr)
    \kappaassoc g^\textit{SW}_\sigma(\xx)
  \left(e^{-\beta\epsilonassoc} - 1\right)}, \label{eq:X}
\end{align}
where
\begin{align}
  n_\text{site}'(\rr) &=
  \begin{cases}
    \frac{4}{\pi\sigma^2} \int n(\rr')\delta(\sigma - |\rr-\rr'|) d\rr' & \text{this work}\\
    4 n_0(\rr) \zeta(\rr) & \text{\hughesetal}
  \end{cases} 
\end{align}
is the density of bonding sites that could bond to the sites~$n_\text{site}(\rr)$, and
\begin{align}
  g^\textit{SW}_\sigma(\xx) &= g^\textit{HS}_\sigma(\xx) +
  \frac{1}{4}\beta\left(\frac{\partial a_1}{\partial \etadisp(\xx)} -
  \frac{\lambdadisp}{3 \etadisp}\frac{\partial a_1}{\partial \lambdadisp}\right)\label{eq:gSW},
\end{align}
where $g^\textit{HS}_\sigma$ is the correlation function evaluated at
contact for a hard-sphere fluid with a square-well dispersion
potential, and $a_1$ and $a_2$ are the two terms in the dispersion
free energy defined below Eq.~\ref{eq:dispersion-term}.  The radial distribution function of the square-well
fluid $g^\textit{SW}_\sigma$ is written as a perturbative correction
to the hard-sphere radial distribution function
$g^\textit{HS}_\sigma$.  The functional of \hughesetal\ uses the
$g_\sigma^\textit{HS}$ from Yu and Wu
\cite{yu2002fmt-dft-inhomogeneous-associating}.  In this work, we use
the $g_\sigma^\textit{HS}$ derived by Schulte~\emph{et~al.}~\cite{schulte2012using}.

As in \hughesetal, we use Clark's five empirical parameters, and fit
the calculated surface tension to experimental surface tension at
ambient conditions by tuning the parameter $\lscale$, which adjusts
the length-scale of the average density used for the dispersion
interaction.  With the improved association term, we find these agree
when $\lscale$ is 0.454, which is an increase from the value of 0.353
found by \hughesetal.  In order to explore further the change made by
the improved association term, we compared the new functional with
that of \hughesetal\ for the two hydrophobic cases of the hard rod and
the hard spherical solute.

\begin{figure}
\begin{center}
\includegraphics[width=3.5in]{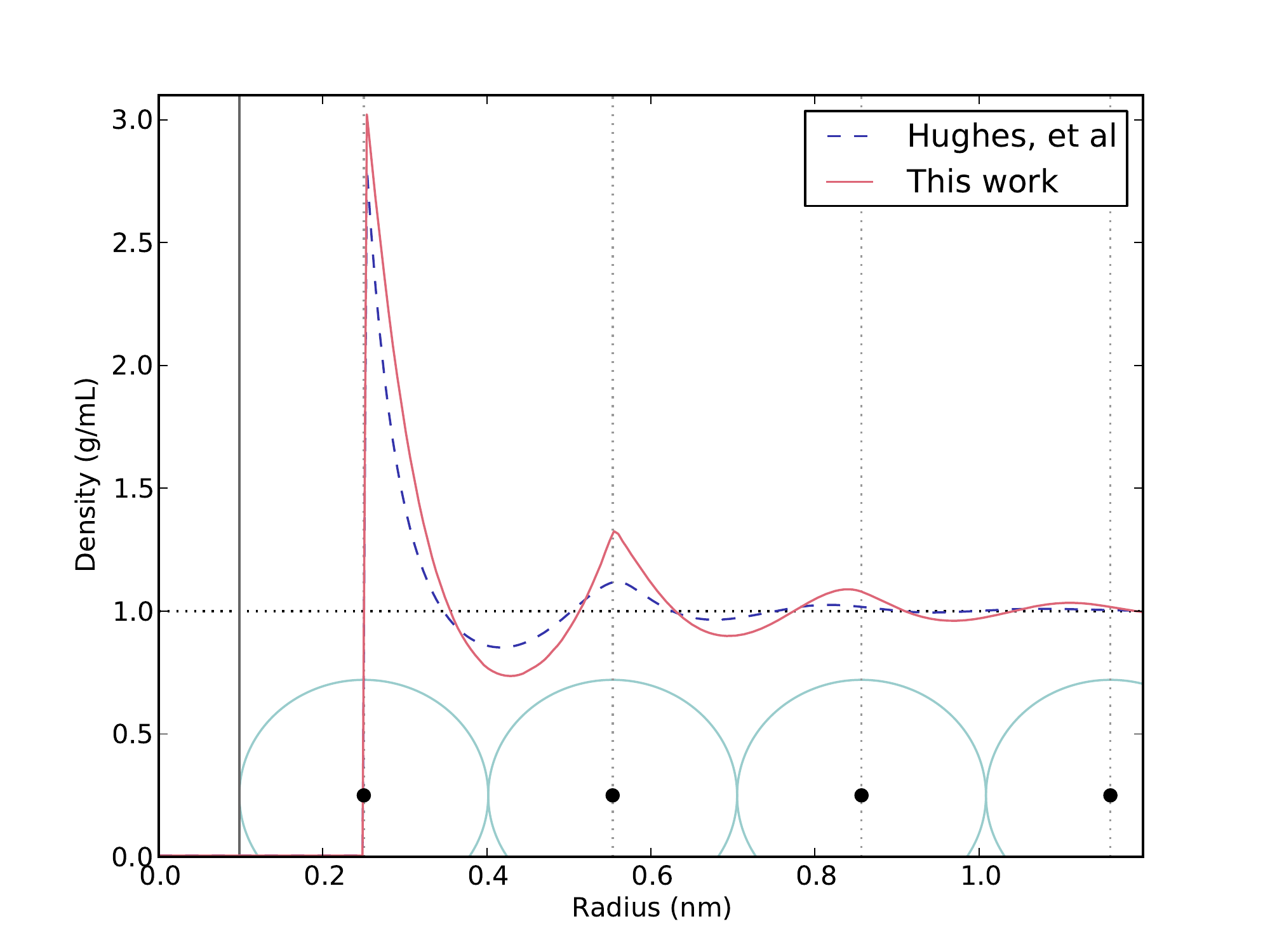}
\end{center}
\caption{ Density profiles for a water around a single hydrophobic rod
  of radius 0.1~nm. The solid red profile is from the functional
  developed in this paper and the dashed blue profile is the result
  from \hughesetal.  For scale, under the profiles
  is a cartoon of a string of hard spheres touching in one
  dimension. The horizontal black dotted line is the bulk density for
  water and the vertical line on the left at 0.1~nm represents the
  rod wall.}
\label{fig:density-single-rod}
\end{figure}

\section{Results}

We will first discuss the case of a single hydrophobic rod immersed in
water. Figure~\ref{fig:density-single-rod} shows the density profile
of water near a rod with radius~1~\AA.  The density computed using the
functional of this paper is qualitatively similar to that from
\hughesetal, with a comparable density at contact---consistent with
having made only a moderate change in the free energy.  The first
density peak near the surface is higher than that from \hughesetal,
and the peak has a kink at the top.  This reflects the improved
accuracy of the $g_\sigma^\textit{HS}$ from \hughesetal, since beyond
the first peak water molecules are unable to touch---or hydrogen bond
to---molecules at the surface of the hard rod. This is illustrated
under the profiles in Figure~\ref{fig:density-single-rod} by a cartoon
of adjacent hard spheres that are increasingly distant from the hard
rod surface.

\begin{figure}
\begin{center}
\includegraphics[width=3.5in]{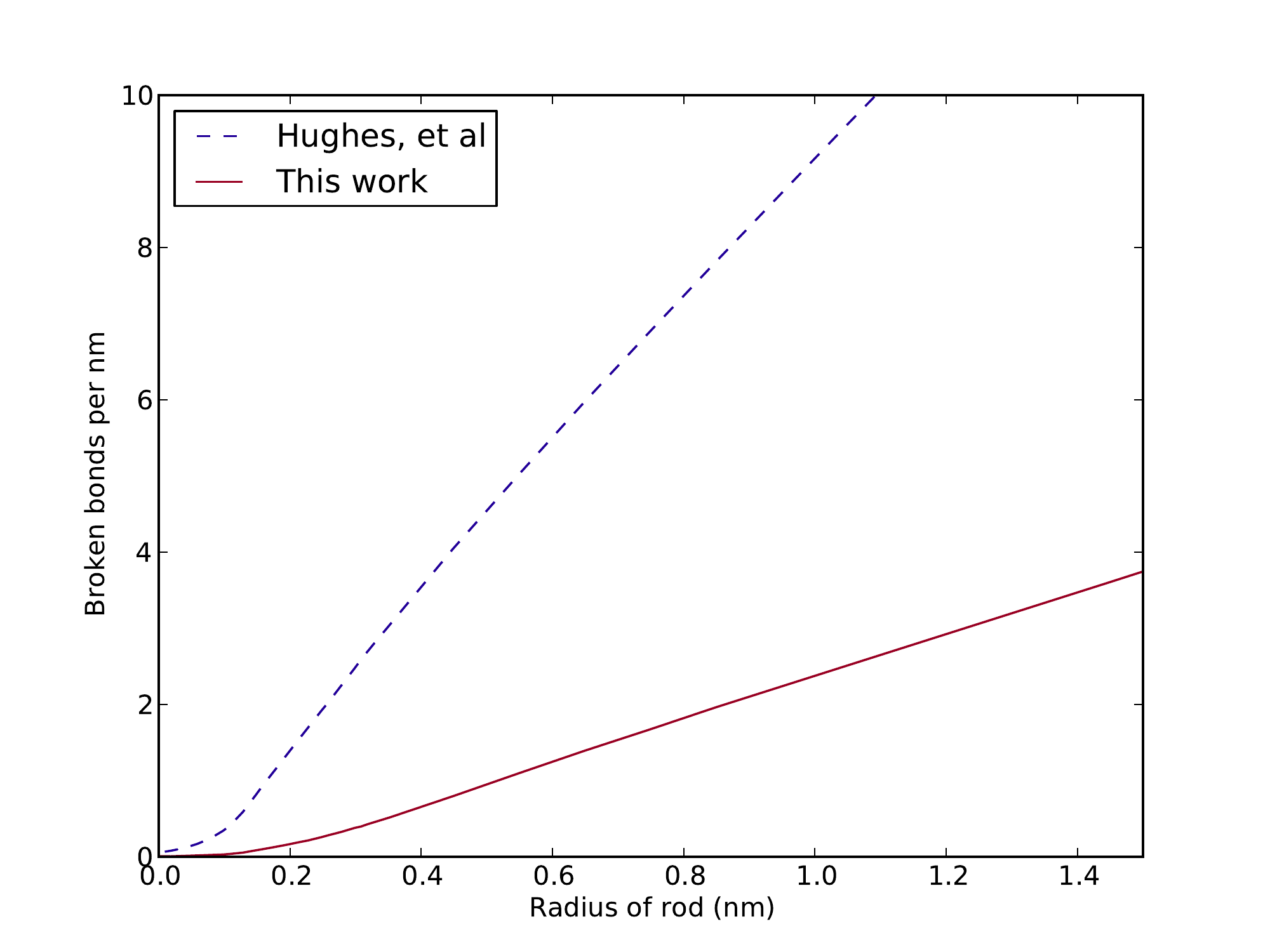}
\end{center}
\caption{Broken hydrogen bonds per nanometer for hydrophobic rods
  immeresed in water.  The solid red line uses the functional developed
  in this paper while the dashed blue line uses the functional from
  \hughesetal. For large enough rods, the graph
  increases linearly for both functionals.}
\label{fig:single-rod-broken-HB}
\end{figure}

In addition to the density, we examine the number of hydrogen bonds
which are broken due to the presence of a hard rod.  We define this
quantity as
\begin{align}
  N_{\text{broken HB}} &= 2 \int (X(\rr) - X_{\text{bulk}})n_{\text{site}}(\rr) d\rr
\end{align}
where $X_{\text{bulk}} = 0.13$ is the fraction of unbonded
association sites in the bulk.  The factor of 2 is chosen to account
for the four association sites per molecule, and the fact that each
broken hydrogen bond must be represented twice---once for each of the
molecules involved.  In Fig.~\ref{fig:single-rod-broken-HB} we show
the number of hydrogen bonds broken by a hard rod per nanometer
length, as predicted by the functional of \hughesetal\ (dashed line)
and this work (solid line), as a function of the radius of the hard
rod.  In each case in the limit of large rods, the number of broken
bonds is proportional to the surface area.  At every radius, the
functional of \hughesetal\ predicts approximately four times as many
broken hydrogen bonds as the improved functional.

\begin{figure}
\begin{center}
\includegraphics[width=3.5in]{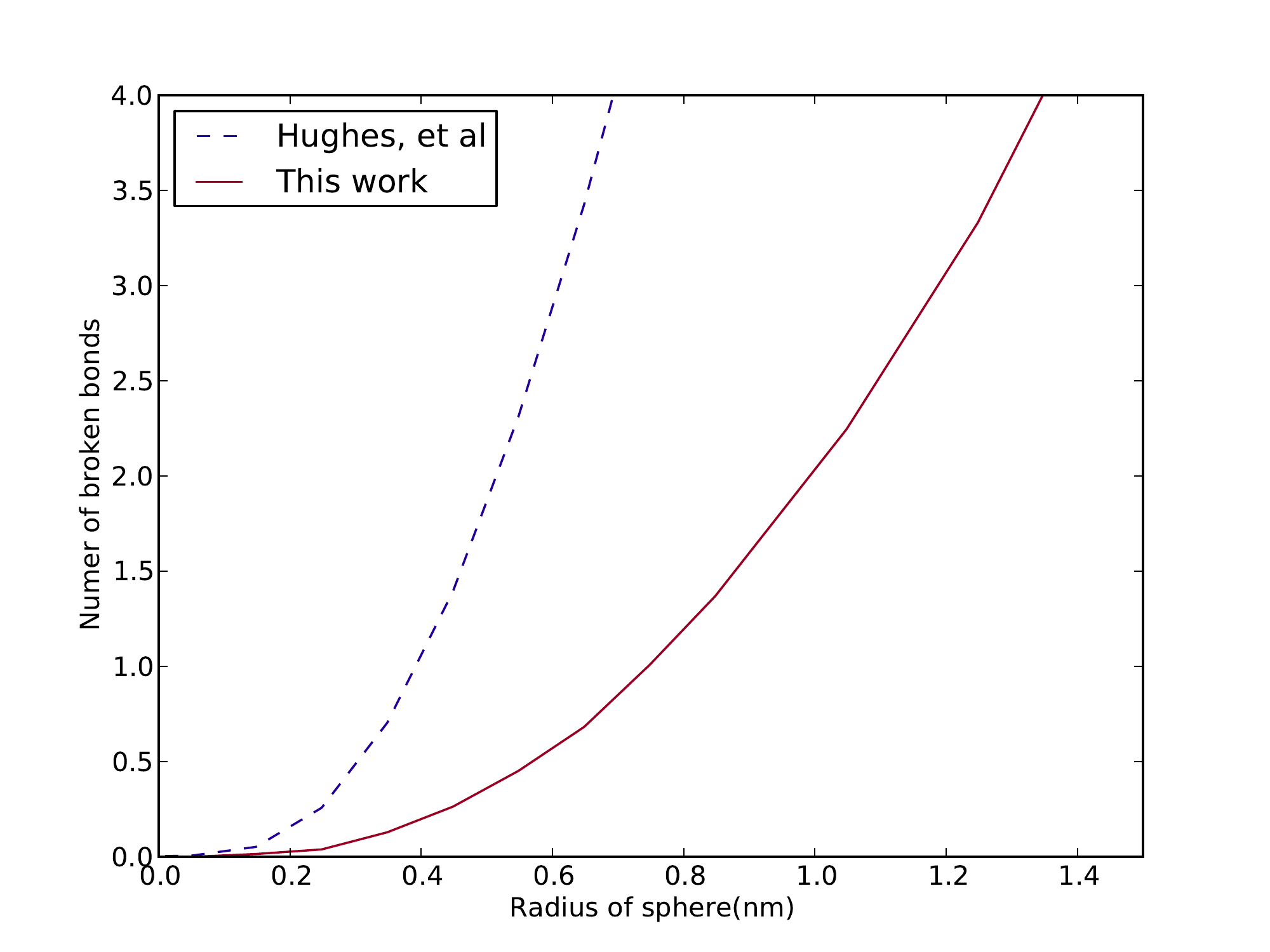}
\end{center}
\caption{Broken hydrogen bonds for hard spheres immeresed in water.
  The solid red line uses our the functional developd in this paper
  while the dashed blue line is from
  \hughesetal.}
\label{fig:spheres-broken-HB}
\end{figure}

A common test case for studying hydrophobic solutes in water is the
hard-sphere solute.  Figure~\ref{fig:spheres-broken-HB} shows results
for the number of broken hydrogen bonds caused by a hard-sphere solute,
as a function of the solute radius.  As in
Fig.~\ref{fig:single-rod-broken-HB}, the number of broken bonds scales
with surface area for large solutes, and the number of broken bonds is
about four times smaller than the number from the functional of
\hughesetal.  For solutes smaller than
3~\AA\ in radius, there is less than a tenth of a hydrogen bond
broken. This is consistent with the well-known fact that small solutes
(unlike large solutes) do not disrupt the hydrogen-bonding network of
water~\cite{chandler2005}.

\section{Conclusion}

We have modified the classical DFT for water developed by
Hughes~\emph{et al.}~\cite{hughes2013classical} with the more accurate
radial distribution function at contact developed by Schulte~\emph{et
  al.}~\cite{schulte2012using}, which affects the predicted hydrogen
bonding between water molecules.  We found that while this
modification has a relatively mild effect on the free energy and
density profiles, it predicts fewer broken hydrogen bonds around
hydrophobic solutes and at aqueous interfaces.

\bibliography{paper}

\end{document}